\documentclass[a4paper,
               keeplastbox,   
               ]{jacow}
\usepackage{pdfpages,multirow,ragged2e} %
\makeatletter%
        \ifboolexpr{bool{xetex}}
         {\renewcommand{\Gin@extensions}{.pdf,%
                            .png,.jpg,.bmp,.pict,.tif,.psd,.mac,.sga,.tga,.gif,%
                            .eps,.ps,%
                            }}{}
\makeatother

\ifboolexpr{bool{xetex} or bool{luatex}} 
 {}                                      
 {\usepackage[utf8]{inputenc}}           

\usepackage[USenglish]{babel}

\begin{document}

\title{Reducing Misinformation in Query Autocompletions\thanks{This 
  study was done while the author worked at Searsia.
  Published at the 2nd International Symposium on Open Search Technology, 12-14 October 2020, CERN, Geneva, Switzerland.}
}

\author{Djoerd Hiemstra,\thanks{djoerd.hiemstra@ru.nl} Radboud University, The Netherlands} 

\maketitle

\begin{abstract}
Query autocompletions help users of search engines to speed up their 
searches by recommending completions of partially typed queries in a 
drop down box. These recommended query autocompletions are usually 
based on large logs of queries that were previously entered by the
search engine's users. Therefore, misinformation entered
-- either accidentally or purposely to manipulate the search
engine -- might end up 
in the search engine's recommendations, potentially harming
organizations, individuals, and groups of people. 
This paper proposes an alternative approach for generating
query autocompletions by extracting anchor texts from a large web 
crawl, without the need to use query logs. 
Our evaluation shows that even though query log autocompletions perform
better for shorter queries, anchor text autocompletions outperform
query log autocompletions for queries of 2 words or more.
\end{abstract}

\section{Introduction}
\begin{figure}[!ht]
  \centering
    \includegraphics[width=\columnwidth]{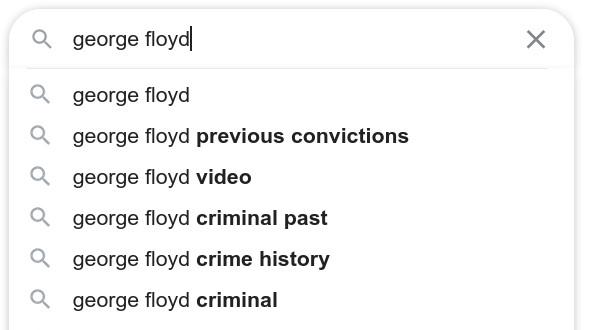}
    \centering
  \caption{Example autocompletions}
    \label{fig:google}
\end{figure}

The brutal killing end of May 2020 by Minneapolis police officers of 
George Floyd, who was already handcuffed, laying face down, 
and did not seem to resist arrest, 
became an immediate target of disinformation on the platforms run 
by Google and Facebook.
Figure \ref{fig:google} shows Google's autocompletions for
George Floyd early June 2020. Although it is hard to proof the deliberate
manipulation of Google's autocompletions in this particular case, we show 
below that autocompletions based on previous user interactions have been 
shown to contain defamatory, racist, sexist and homophobic information, 
and there is increasing evidence that autocompletions are an easy target 
for spreading fake news and propaganda.

Search engines suggest completions of partially typed queries to help users 
speed up their searches, for instance by showing the suggested completions 
in a drop down box. These {\em query autocompletions} enable the user to 
search faster, searching for long queries with relatively few key strokes. 
Jakobsson \cite{Jakobsson1986} showed that for a library information system, 
users are able to identify items using as little as 4.3 characters on average. 
Query autocompletions are now widely used, either with a drop down box or as 
instant results \cite{BastWeber2006}. 
While Jakobsson \cite{Jakobsson1986} used
the titles of documents as completions of user queries, web search engines 
today generally use large logs of queries submitted previously by their users. 
Using previous queries seems a common-sense choice: The best way to predict a 
query? Use previous queries! Several scientific studies use the AOL query log 
provided by Pass et al.\ \cite{Pass2006} to show that query autocompletion 
algorithms using query logs are effective 
\cite{BarYossefKraus2011, ShokouhiRadinsky2012, WhitingJose2014, MitraCraswell2015, Cai2016}. 
However, query autocompletion algorithms that are based on query logs are 
problematic in two important ways: 
\begin{enumerate}
\item They return offensive and damaging results; 
\item They suffer from a destructive feedback loop.  
\end{enumerate}
We discuss these two problems in the following sections. 

\subsection{Offensive queries and misinformation in query autocompletions}

Query autocompletion based on actual user queries may return offensive results,
and there are several examples where offensive autocompletions
hurt organizations or individuals. For instance in 2010,
a French appeals court ordered Google to remove the word ``arnaque'', 
which translates roughly as ``scam'', from appearing in Google's 
autocompletions for the CNFDI, the Centre National Priv\'e de Formation 
a Distance \cite{McGee2010}. 
Google's defense argued that its tool is based on an algorithm applied to 
actual search queries: It was users that searched for ``cnfdi arnaque'' that 
caused the algorithm to select offensive suggestions. The court however ruled 
that Google is responsible for the suggestions that it generates, and that
Google should remove misinformation that is based on user-generated input of
its search engine. 
Google lost 
a similar lawsuit in Italy, where queries for an unnamed plaintiff's name were 
presented with autocomplete suggestions including ``truffatore'' (``con man'') 
and ``truffa'' (``fraud'') \cite{Meyer2011}. In another similar law suite,
German's former first lady Bettina Wulff sued Google for defamation when 
queries for her name completed with terms like ``escort'' and ``prostitute'' 
\cite{Lardinois2012}. 
In yet another lawsuit in Japan, Google was ordered to disable autocomplete 
results for an unidentified man who could not find a job because a search for
 his name linked him with crimes he was not involved with \cite{BBC2012}. 

Google has since updated its autocompletion results by filtering offensive 
completions for person names, no matter who the person is \cite{Yehoshua2016}. 
But controversies over query autocompletions remain. A study by Baker and Potts
\cite{BakerPotts2013} highlights that autocompletions produce suggested 
terms which can be viewed as racist, sexist or homophobic. Baker and Pots 
analyzed the results of over 2,500 query prefixes and concluded that 
completion algorithms inadvertently help to perpetuate negative stereotypes 
of certain identity groups. A study by Ray and Ayalon \cite{Roy2019} 
suggests that 
Google plays an important role in the spread of age and gender stereotypes via 
its autocomplete algorithm. And despite Google intention to filter autocompletions for person names, 

There is increasing evidence that autocompletions play an important role in 
spreading fake news and propaganda. Query suggestions actively direct users 
to fake content on the web, even when they are not looking for it 
\cite{Roberts2016}. Examples include completions like 
``Did the holocaust happen'', which if selected, returned as its top result 
a link to the neo-Nazi site stormfront.org \cite{Cadwalladr2016}. 
Bad publicity will usually persuade Google to remove such autocompletions. 
In 2016, Google announced it removed ``are Jews evil'' from its autocompletions,
but many similar offensive completions were still suggested two years later
\cite{Lapowsky2018}. Removing such completions is important, because they
led people to search for offensive results that otherwise would not have.
Stephens-Davidowitz \cite{Davidowitz2019} showed that in the 12 months 
following Google's removal of ``are Jews evil'', approximately 10\% fewer 
such questions were asked compared to the 12 months before the removal.

The examples show that query autocompletions can be harmful if they are 
based on searches by previous users. Harmful completions are suggested when 
ordinary users seek to expose or confirm rumors and conspiracy theories. 
Furthermore, there are indications that harmful query suggestions increasingly 
result from  computational propaganda, i.e., organizations use bots to game 
search engines and social networks \cite{ShoreyHoward2016}.
It is not hard to manipulate search autocompletions, as shown by Want 
et al.\ \cite{Wang2018}, who revealed hundreds of thousands of manipulated 
terms that are promoted through major search engines.

\subsection{A destructive feedback loop}

Misinformation and morally unacceptable query completions are not only 
introduced by the searches of previous users, they are also mutually 
reinforced by the search engine and its users. 
When a query autocompletion algorithm suggests morally unacceptable 
queries, users are likely to select those, even if the users are only 
confused or stunned by the suggestion. But how does the search engine 
ever learn it was wrong? It might not ever. As soon as the system 
determined that some queries are recommended; they are more of
them selected by users, which in turn makes the queries end up in the 
training data that the search engine uses to train it's future query 
autocompletion algorithms. Such a destructive feedback loop is one of 
the features of a \emph{Weapon of Math Destruction}, a term coined by 
O'Neil \cite{ONeil2016} to describe harmful statistical models. 

O'Neil sums up three elements of a Weapon of Math Destruction: Damage, 
Opacity, and Scale. Indeed, the \emph{damage} caused by query 
autocompletion algorithms is extensively discussed in the previous 
section. Query autocompletion algorithms are \emph{opaque} because they 
are based on the proprietary, previous searches known only by the search 
engine. If run by a search engine that has a big market share, the query 
completion algorithm also \emph{scales} to a large number of users. Query 
autocompletions of a search engine with a majority market share in a 
country might substantially alter the opinion of the country's citizens, 
for instance, a substantial number of people will start
to doubt whether the holocaust really happened.

\subsection{Structure of the paper}
This paper is structured as follows. In Section \ref{sec:method}, we 
describe a simple but powerful approach that trains query autocompletions
using the content that is indexed by the search engine by extracting
anchor texts from a large web crawl. Section \ref{sec:results} compares
these content-based query autocompletions to collaborative query autocompletions
 based on query logs. Section \ref{sec:conclusion} concludes the paper.

\section{Content-based autocompletions}
\label{sec:method}

It is instructive to view a query autocompletion algorithm as a recommender 
system, that is, the search engine recommends queries based on some input. 
Recommender systems are usually classified into two categories based on how 
recommendations are made \cite{AdomaviciusTuzhilin2005}:
\begin{enumerate} 
\item Collaborative recommendations, and 
\item Content-based recommendations. 
\end{enumerate}
Collaborative query autocompletions are based on similarities between users:
``People that typed this prefix often searched for: \ldots''. 
Content-based query autocompletions are based on similarities with the content:
``Web pages that contain this prefix are often about: \ldots''. 

Until now, we only discussed collaborative query autocompletion algorithms. 
What would a content-based query autocompletion algorithm look like? 
Bhatia et al.\ \cite{Bhatia2011} proposed a system that generates 
autocompletions by using all $N$-grams of order 1, 2 and 3 (that is single 
words, word pairs, and word triples) from the documents. They tested their 
content-based autocompletions on newspaper data and on data from 
ubuntuforums.org. Instead of $N$-gram models from all text, Kraft and Zien
\cite{KraftZien2004} built models for query reformulation solely from the 
anchor texts, the clickable texts from hyperlinks in web pages. Interestingly, 
Dang and Croft \cite{DangCroft2010} argue that anchor text can be an 
effective substitute for query logs. They studied the use of anchor texts 
for a range of query reformulation techniques, including query-based 
stemming and query reformulation, treating the anchor texts as if it were 
a query log.

Inspired by research of Bhatia et al.\ \cite{Bhatia2011}, 
Kraft and Zien \cite{KraftZien2004}, and Dang and Croft \cite{DangCroft2010}, 
we obtain query autocompletions from the anchor texts of web pages,
and test how well these autocompletions predict full queries from
a large query log of a web search engine, given a query prefix.

\section{Collaborative vs.\ Content-Based autocompletions}
\label{sec:results}

In this section, we answer the question: Are query suggestions from anchor 
texts any good compared to query suggestions from query logs? To evaluate  
this, we used the query log of Pass et al.\ \cite{Pass2006}, 
which contains 20 million queries submitted by about 650,000 users to the 
AOL search engine between March and May in 2006. Following the recent 
query autocompletion experiments by Cai et al.\ \cite{Cai2016}, 
we used queries submitted before 8 May 2006 as training queries and queries 
submitted afterwards as test queries. We removed queries containing URL 
substrings (`http:', `https:', `www.', `.com', `.net', `.org', and `.edu') 
from both the training and the test queries. We did not further filter the 
data (Cai et al.\ \cite{Cai2016} only kept queries appearing in both 
partitions). Because we are not interested in personalization, we 
put 99\% of the users in the training data, and the remaining 1\% of the users 
in the test data. This leaves more than 3.3 million unique training queries 
and 952 queries for testing the system. For every test query, we used 10 
different prefixes as input to: 1 to 5 characters, and 1 to 5 words. If 
the query has less than 5 characters or less than 5 words, we take the full 
query as input. For each prefix we measured the mean reciprocal rank (MRR) of the 
position for which our approaches return the full test query. The MRR 
is calculated as one divided by the position of the correct 
result, so it will be 1 if the correct result is returned first, 0.5 if it 
is returned second, etc. We also measured the average number of results 
returned for each prefix. The results of autocompletions using the 3.3 
million unique queries from the training data are presented in 
Table~\ref{tab:logbased}.

\begin{table}[!ht]
    \centering
    \caption{Quality of autocompletions using query logs}
    \begin{tabular}{lrr}
    \toprule
\textbf{Prefix} &  \textbf{MRR} & \textbf{Returned} \\
    \midrule
1 char & 0.026 &  10.00 \\
2 char & 0.072 &  10.00 \\
3 char & 0.135 &   9.99 \\
4 char & 0.181 &   9.71 \\
5 char & 0.227 &   9.28 \\
1 word & 0.271 &   8.15 \\
2 word & 0.354 &   4.40 \\
3 word & 0.365 &   3.30 \\
4 word & 0.365 &   3.06 \\
5 word & 0.366 &   3.04 \\
    \bottomrule
    \end{tabular}
    \label{tab:logbased}
\end{table}

Table \ref{tab:logbased} shows that after typing 3 characters the MRR is 
0.135, so the correct suggestion is on average available in the top 8 
results (1/8 = 0.125). After typing 1 word, the MRR is 0.271, i.e., on 
average the correct suggestion is available in the top 4 results. 

For anchor text completions we ideally would need a large web crawl 
from 2006 (the year of the query log). In absence of such a crawl, we used 
data from ClueWeb09, a web crawl of more than 1 billion 
pages crawled in January and February 2009 by Callan and Hoy \cite{Clueweb09} 
at Carnegie Mellon University. Anchor texts for the English pages in this 
collection (about 0.5 billion pages) are readily available 
\cite{HiemstraHauff2010}, 
so we do not actually need to process the ClueWeb09 web pages themselves. 
Anchor texts with  separators  (`.', `?',`!', `$|$', `-' or `;') followed by a 
space were split in multiple strings. Text in braces `()', `\{\}', `[]' was removed 
from the strings. We processed the anchor texts by retaining only suggestions 
that occur at least 15 times. This resulted in 46 million unique suggestions. 
Performance of the ClueWeb09 anchor text suggestions is presented in 
Table~\ref{tab:anchorbased}.

\begin{table}[!ht]
    \centering
    \caption{Quality of autocompletions using anchor texts}
    \begin{tabular}{lrr}
    \toprule
\textbf{Prefix} &  \textbf{MRR} & \textbf{Returned} \\
    \midrule
1 char & 0.006 &  10.00 \\
2 char & 0.025 &  10.00 \\
3 char & 0.066 &  10.00 \\
4 char & 0.123 &   9.92 \\
5 char & 0.180 &   9.71 \\
1 word & 0.251 &   8.62 \\
2 word & 0.415 &   5.42 \\
3 word & 0.440 &   3.95 \\
4 word & 0.443 &   3.67 \\
5 word & 0.443 &   3.64 \\
    \bottomrule
    \end{tabular}
    \label{tab:anchorbased}
\end{table}

Table~\ref{tab:anchorbased} shows that for queries of 2 words or more (the 
average query length in the test data is 2.6), anchor text autocompletions 
perform better than query log autocompletions, up till an MRR of 0.443 
(0.366 for query logs). Query log autocompletions perform better for shorter 
queries: Anchor text autocompletions need about one character more than 
query log autocompletions to achieve a similar MRR for the first 5 characters.
The source code for running the experiment is available from 
\url{https://github.com/searsia/searsiasuggest}.

\section{Conclusion}
\label{sec:conclusion}

Query autocompletions based on anchor text from web pages perform remarkably 
well. For queries of more than one word, they outperform autocompletions that 
are based on over two months of query log data. Simply extracting all anchor 
texts is really only a first attempt to get well-performing autocompletions 
from web content. Ideas to improve suggestions are: Using linguistic knowledge 
to get suggestions from all web page text (for instance using the Stanford 
CoreNLP tools \cite{StanfordCoreNLP}), using web knowledge like PageRank 
scores and Spam scores\footnote{PageRank scores and Spam scores are also 
available for ClueWeb09 \cite{Clueweb09}.}
to improve the quality of suggestions, and reranking of suggestions by their 
``query-ness'' using machine learning. 

Future work should follow a user-centered evaluation, using 
ethical instruments of analysis, to better measure the usefulness of
autocompletions. This includes measuring if can suggest a query that is
better than the user's intended query, measuring the actual amount of 
misinformation in autcompletions (links can also be manipulated, using 
so-called Google bombing), as well as their timeliness (updating from 
hyperlinks might be slower than updating from queries).

\paragraph{Acknowledgments}
I am grateful to the Vietsch Foundation and NLnet Foundation 
for funding the work presented in this paper, which
was done at Searsia (\url{http://searsia.org}).

\bibliographystyle{plain}



\end{document}